# The effect of direct electron beam patterning on the water uptake and ionic conductivity of Nafion thin films

*Ky V. Nguyen, Jan G. Gluschke, A. Bernardus Mostert, Andrew Nelson, Gregory Burwell, Roman W. Lyttleton, Hamish Cavaye, Rebecca J.L. Welbourn, Jakob Seidl, Maxime Lagier, Marta Sanchez Miranda, James D. McGettrick, Trystan Watson, Paul Meredith, and Adam P. Micolich\**

K. V. Nguyen, J. G. Gluschke, J. Seidl, M. Lagier, M. Sanchez Miranda, A. P. Micolich
School of Physics, University of New South Wales, Sydney NSW 2052, Australia
E-mail: adam.micolich@nanoelectronics.physics.unsw.edu.au

A.B. Mostert, G. Burwell
Centre for Integrative Semiconductor Materials, Department of Physics, Swansea University
Bay Campus, Fabian Way, Swansea SA1 8EN

A. Nelson
Australian Nuclear Science and Technology Organisation, Lucas Heights NSW 2234, Australia

R. W. Lyttleton
Solid State Physics and NanoLund, Lund University, Box 118, 22100 Lund, Sweden

H. Cavaye, R. J. L. Welbourn
Isis Neutron and Muon Source, Rutherford Appleton Laboratory, Science and Technology Facilities Council, Didcot OX11 0QX, U.K.

J. D. McGettrick, T. Watson
SPECIFIC, College of Engineering, Bay Campus, Swansea University, Fabian Way, Crymlyn Burrows, Swansea SA1 8EN, Wales, U.K.

P. Meredith
Centre for Integrative Semiconductor Materials, Department of Physics, Swansea University
Bay Campus, Fabian Way, Swansea SA1 8EN
School of Mathematics and Physics, University of Queensland, Brisbane QLD 4072, Australia

**Abstract:** We report the effect of electron-beam patterning on the water uptake and ionic conductivity of Nafion films using a combination of x-ray photoelectron spectroscopy, quartz crystal microbalance studies, neutron reflectometry, and AC impedance spectroscopy. The aim was to more fully characterize the nature of the nanoscale patterned Nafion structures recently used as a key element in novel ion-to-electron transducers by Gluschke *et al.* To enable these studies, we develop the electron beam patterning process for large areas, achieving patterning speeds approaching 1 cm$^2$/hr, and patterned areas as large as 7 cm$^2$ for the neutron reflectometry studies. We ultimately show that electron-beam patterning affects both the water uptake and the ionic conductivity, depending on film thickness. We see Type-II adsorption isotherm behaviour for all films. For thick films (~230 nm), we see a strong reduction in water uptake with electron-beam patterning. In contrast, for thin films (~30 nm), electron-beam patterning enhances water uptake. Notably, we find that for either thickness the reduction in ionic conductivity arising from electron-beam patterning is kept to less than an order of magnitude. We propose mechanisms for the observed behaviour based on the known complex morphology of Nafion films to motivate future studies of electron-beam processed Nafion.

## 1. Introduction

Nafion is a perfluorinated polymer with sulfonated pendant groups first developed by Walther Grot at DuPont in the 1970s.[1] Its dominant practical use is as a proton exchange membrane in fuel cells[2,3] owing to an exceptionally high protonic conductivity,[4] which arises from a unique microstructure of water nanochannels[5–7] threading through the polymer matrix. Recently Nafion has emerged as a material with significant potential for bioelectronics and neuromorphic computing applications. For example, Josberger et al.[8] demonstrated synaptic-like short-term depression behaviour and memory effects in two-terminal devices featuring a Nafion film spanning a pair of hydrogenated palladium contacts. van de Burgt et al.[9] and Fuller et al.[10] have reported low-voltage artificial synapse structures featuring a Nafion film sandwiched between PEDOT:PSS conducting polymer electrodes, which can be deployed in ionic floating-gate memory arrays for scalable neuromorphic computing. Nafion films have featured in other applications including as an iontronic element in electronic skin interfaces,[11] as a memristor in an artificial touch-sensory nerve system,[12] a controllable sodium ion reservoir,[13] and as ion-conducting membrane structures for artificial photosynthesis devices.[14,15]

In each instance above, the Nafion was unstructured and used either as a) a thin-film that was drop-cast[8,10,14] or spin-coated[11–13] onto a substrate, or b) a Nafion-soaked cellulose tissue[8] or a freestanding Nafion-117 membrane sandwiched between two layers.[9] The ability to lithographically pattern a thin Nafion film at the micro- and nanoscale is important to the further development of Nafion for device applications. We recently demonstrated a first step in this direction by using electron-beam lithography to generate nanoscale Nafion structures with linewidths down to 125 nm, which were subsequently used as ionic gating structures for III-V nanowire transistors.[16] We chose electron-beam lithography, and continue to use it here despite the large patterned areas required for this particular work, for three reasons. Firstly, solubility contrast cannot be obtained in pure Nafion using traditional UV lithography, which would be the obvious alternative option. Secondly, other alternatives are either not viable, e.g., inkjet printing has insufficient resolution for use with nanowires, or carry detrimental effects, e.g., ion-beam lithography, which causes significant damage to the material and substrate. Thirdly, electron-beam lithography has demonstrated efficacy producing solubility contrast, including both positive and negative contrast, for pure polymer films, e.g., polymethylmethacrylate[17] and polyethylene oxide[18,19], respectively. Ultimately, we produced microscale integrated complementary ion-to-electron signal transduction circuits with nanoscale Nafion features that gave DC gain exceeding 5 and frequency response up to 2 kHz.[16] The frequency response compared well with PEDOT:PSS-based organic electrochemical transistors, where response in the kHz is the current state-of-the-art.[20]

The notable aspect here is that Nafion has a negative contrast under electron-beam patterning, which confers a major advantage for deployment of Nafion in micro- and nanoscale device applications because only the regions that remain in the device after development need to be exposed, and their cumulative area is small at $\mu m^2$ to $mm^2$ per chip.[8,10,16,18,19] This makes patterning a time-effective process with write times below one second for a 1 $\mu m^2$ structure.[16] We show in this work that even considerably larger cumulative areas of Nafion can be patterned efficiently (7 cm$^2$ at ~0.86 cm$^2$/hr) using electron-beam processing with some careful optimisation of the patterning process, opening a path to scalable use of electron-beam patterned Nafion in device applications. A further advantage of electron-beam lithography is that the energy intensity delivered at the resist by the electron-beam is of order 10 μJ/cm$^2$, several orders of magnitude lower than needed in photolithography for modern photoresists, e.g., 50-150 mJ/cm$^2$ for ECI-3012 (positive) or AZ nLOF2020 (negative). This implies that the electron-beam approach would potentially lead to less overall damage to the exposed material relative to UV photolithography.

Our approach to nanoscale patterning of Nafion films raises interesting fundamental questions about the material that remains after patterning: How does our electron-beam processing affect the chemistry, water absorption, and ultimately the ionic conductivity of the Nafion relative to the equivalent unpatterned 'pristine' material? The latter is particularly relevant in the context of bioelectronics and neuromorphic computing, where Nafion is selected primarily for its superior ion-transport properties.

In this paper, we use x-ray photoelectron spectroscopy (XPS), quartz crystal microbalance (QCM) measurements, neutron reflectometry (NR) and AC impedance spectroscopy (ACIS) to study how key materials properties differ between pristine and electron-beam processed Nafion films. We perform our materials characterization on two different target film thicknesses of Nafion: 30 nm and 230 nm. The 230 nm films are what were used for the device structures in Gluschke *et al.*[16] They are expected to exhibit properties similar to the well-studied Nafion membranes of earlier literature.[21] The 30 nm films enable higher patterning resolution,[16] and are of interest because of a well-known change in the hydration behaviour in Nafion for film thicknesses less than 60 nm.[21–24] In keeping with Kusoglu *et al.*,[23] we will hereafter refer to our 230 nm films as bulk-like or thick films and our 30 nm films as thin films. We also note in advance that the actual final measured thickness of our films can differ from the target thickness for reasons of real thickness variability arising from the processing and measurement uncertainty. We discuss this further in the results and methods sections and note that general usage of 230 nm and 30 nm defaults to target thickness for the discussion hereafter.

## 2. Results and discussion

A significant challenge with XPS, QCM, ACIS, and NR characterisations of an electron-beam patterned material is that large, patterned areas ($cm^2$) are required. This can impose significant costs if the electron-beam processing is not optimised for large area exposures. Since the electron-beam processing parameters have a flow-on effect on materials properties, any optimisation needs to be considered before, rather than after, the materials characterisation. To accommodate this, we commence with a discussion of the factors involved in optimising the electron-beam patterning process. We then outline the resulting key electron-beam processing used for the material characterisation samples in Section 2.2.1 before considering the characterisation results.

### 2.1. Effects of key parameters in the electron-beam patterning process

**Figure 1**a-c shows an illustration of the direct electron-beam patterning of Nafion. First, the Nafion film was spin-coated on a substrate from a 5% suspension of Nafion-117 (Fig. 1a). The 30 nm films were obtained with the Nafion-117 suspension diluted 1:4 in ethanol. Select areas of the Nafion film were exposed to an electron-beam in a Raith 150-Two (8 $\mu C/cm^2$ dose at 10 keV) or Raith Voyager (20 $\mu C/cm^2$ dose at 50 keV) electron-beam lithography (EBL) system (Fig. 1b). The Raith Voyager system was only used on the largest area exposures due to its higher patterning speed (see Methods). The substrate was then submerged in a 1:1 mixture of 2-propanol and acetone for 60 s. The 'developer' solution dissolves all unexposed Nafion, and only the electron-beam exposed Nafion remains thereafter (Fig. 1c).

This patterning approach is ideal for device applications where nanoscale Nafion structures are required because only the small active regions (~$\mu m^2$) need exposure. A challenge here, where we perform materials characterisations by XPS, QCM, ACIS and NR, is that we require significantly larger patterned films with a total patterned area up to several cm$^2$. The difficulty is that EBL involves raster-scanning of a focussed electron-beam with a fixed current and so the required exposure time scales linearly with the required pattern area. Using exposure parameters optimized for nanoscale device applications[16] results in exposure times of order 100 hours per cm$^2$ of pattern area. This is clearly impractical for the study intended here. Fortunately, there are routes to optimising exposure for larger areas since positional accuracy, resolution and exposure-edge sharpness requirements are substantially reduced. The most obvious initial optimisation for large area is to increase the beam current by opening the electron-beam column aperture, but this only helps up to a certain point. The serpentine raster in an EBL system is achieved via a set of 'dwell points' separated by a settable fixed distance of order tens to hundreds of nanometres. The time that the beam rests at each dwell point ('dwell time') is set by the areal charge dose and beam current, and if the current is high, the dwell time becomes very short. The required beam speed, defined as dwell point spacing divided by dwell time, can then exceed what can be driven by the EBL system's pattern generator electronics. A solution is to increase the dwell point spacing. This increases the dwell time, minimising the beam speed for a given beam current, but has a natural limit as we show later. The total patterning time is not just set by the beam current, dose and pattern area, there is an associated overhead from the pattern generator that can also be optimised. The 'settling time' can be reduced by running the EBL system in 'meander' mode rather than 'line' mode. Additionally, increasing the write-field size cuts down the time associated with stage movement. Ultimately, these optimisations reduce the patterning resolution and precision, which although crucial for ultrafine structures, can be acceptably sacrificed for speed for larger area exposures. Obtaining optimal scalability in future devices with patterned Nafion would likely involve 'hybrid writes', where different structure scales are written with separately optimised settings.

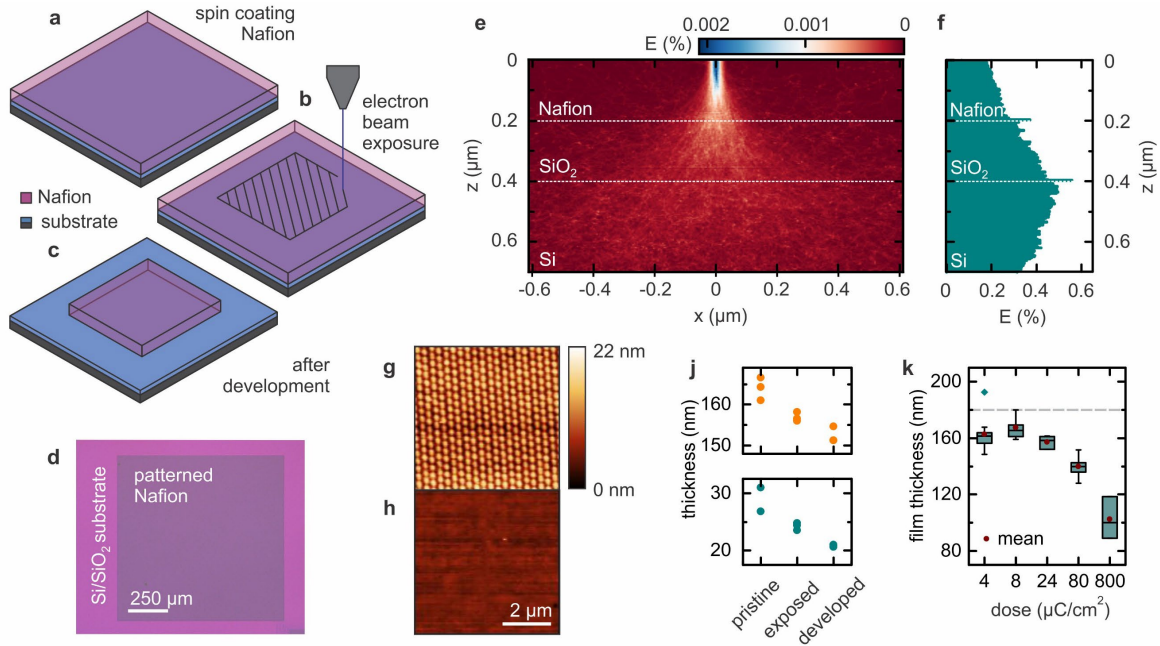

**Figure 1.** (a-c) Illustration of the Nafion patterning process. (a) Nafion film was spin coated on a substrate. (b) Select areas of the film were exposed with an electron-beam. (c) The sample was developed in a solvent mixture, dissolving the unexposed Nafion while the exposed Nafion remains. (d) Optical microscopy of a 1 mm$^2$ patterned Nafion film on a Si/SiO$_2$ substrate. (e) Monte Carlo simulation of the electron-beam exposure of a Nafion/SiO$_2$/Si sample. The heat map shows the percentage of total energy $E$ deposited in a 4.9 × 3.2 nm$^2$ pixel in the *x-z* spatial plane. (f) Percentage of total energy deposited by depth *z* per 3.2 nm slice. (g/h) Atomic Force Microscopy (AFM) surface scan of a Nafion film electron-beam patterned with (g) a focussed beam and (h) a defocussed beam. A 300 nm dwell-point spacing was used in each case. (j) Thickness measurement by ellipsometry of films with target thickness 230 nm (orange) and 30 nm (cyan) before electron-beam exposure (pristine), after electron-beam exposure (exposed), and after development (developed). The measurement was repeated three times at different positions for each data-point. (k) Film thickness of an electron-beam patterned Nafion film after development as a function of area dose. The grey dashed line indicates the thickness of the film prior to patterning as determined by ellipsometry.

A consideration of the spatial energy distribution deposited by the electron-beam is vital when varying exposure parameters. Figure 1e shows the percentage of energy deposited in a ~4.9 × 3.2 $nm^2$ cross section of a 200 nm thick Nafion film on a Si substrate with a native oxide surface for a 10 keV electron-beam with a 25 nm beam-spot radius at the Nafion surface. The data was obtained from a Monte Carlo simulation of 1000 electrons using the CASINO v2.5.1.0 software.[25] The beam diverges as it penetrates the material and some electrons scatter from the Nafion/$SiO_2$ interface. This leads to an increasingly homogeneous energy distribution across the beam width for increasing penetration depth $z$ into the Nafion film. The percentage of energy deposited also increases with $z$, as shown in Fig. 1f. Consequently, portions between dwell points near the Nafion surface may receive insufficient dose for patterning. This can generate periodic surface modulations of order 20 nm in height in the patterned films, as observed in Fig. 1g where a 300 nm dwell point spacing was used. This may have practical use if a non-flat surface is warranted, e.g., the Nafion-catalyst interface in fuel-cell applications[26,27] or brain-material interfaces,[28] since Nafion is biocompatible.[29] Additional surface scans of films patterned with different dwell point spacings are provided in the Supporting Information as Figure S1. Exposure uniformity can be significantly improved by defocussing the electron-beam. Figure 1h shows a surface scan of the same Nafion film patterned using an unfocussed beam instead, giving a surface that is significantly smoother than in Fig. 1g. For an unfocussed beam we achieve a root-mean-square (rms) surface roughness of ~0.9 nm for films patterned with 300 nm dwell point spacing and ~1.2 nm for 80 nm dwell point spacing. This is comparable to the rms surface roughness of ~0.7 nm for a pristine spin-coated Nafion film.

The electron-beam patterning process also affects the overall Nafion film thickness. We find that patterned Nafion is typically 5-10 nm thinner than the pristine film it was patterned from, even for uniform exposures, i.e., defocussed beam or small dwell-point spacing and using our default dose of 8 $\mu C/cm^2$. Figure 1j shows ellipsometry measurements of film thickness before exposure, after exposure, and after development for films with 230 nm and 30 nm target thickness. The data suggests thickness loss occurs in two stages, the first during exposure and the second during development, with a loss of approximately 5 nm at each stage. We attribute the development loss to the removal of surface Nafion that received insufficient dose to become insoluble. The origin of the exposure loss needs more careful consideration. A challenge with ellipsometry measurements is that we cannot rule out small changes in optical properties, e.g., refractive index, absorption, that arise during exposure[16] and may lead to systematic error.

To confirm/validate these thickness changes, we performed atomic force microscopy (AFM) studies. Figure 1k shows the measured thickness of electron-beam patterned regions with different exposure dose, produced in a common film with 230 nm target thickness after a common development step. A similar graph for data obtained before development is shown in Fig. S2. The grey dashed line in Fig. 1k is the actual thickness before patterning as determined by ellipsometry. The underpinning AFM scan for Fig. 1k is given in Fig. S3. The measured thickness in Fig. 1k initially increases with dose for doses up to ~10 µC/cm$^2$ and thereafter decreases by as much as 65 nm for a dose of 800 µC/cm$^2$. The local thinning with high doses is consistent with observations by Steinbach *et al.*,[30] who used a 50 keV electron beam and 800 µC/cm$^2$ dose to create micron-scale channels in Nafion membrane. Notably, the thickness reduction at higher doses observed in Fig. 1k is not linear, and we found that it saturates at ~100 nm at around 5000 µC/cm$^2$ (see Fig. S2). This suggests that the Nafion is compressed rather than being removed, e.g., by ablation. A possible mechanism for this behaviour is as follows. First, the vacuum during the electron-beam exposure removes water, de-swelling the Nafion film.[24,31] The electron-beam exposure then generates crosslinking (see XPS discussion in Section 2.2.2). This locks some of the compressed polymer chains into place, preventing the film from swelling to its original size under a return to ambient conditions. The dependence of film thickness on exposure dose is interesting because it opens opportunities for three-dimensional sculpting of Nafion nanostructures beyond the basic two-dimensional patterning demonstrated in earlier work.[16] We do not propose that our approach is the ideal way to achieve surface textures in Nafion – there are other well-known methods for surface texturing of polymer films that are faster and cheaper – we simply point out that it could be obtained as part of our broader nanoscale patterning process simply by appropriately tuning the patterning parameters.

## 2.2. Materials characterization of electron-beam patterned Nafion films

We now shift our focus to how electron-beam patterning affects other properties of the Nafion films. This requires us to choose a set of exposure parameter values for the electron-beam exposure, which is guided in part by our findings in Section 2.1, but also heavily driven by the exposure parameters used in our previous work on Nafion-gated nanowire ion-to-electron transducers.[16]

## 2.2.1. Exposure, preparation, and handling details for materials characterisation samples

For the XPS, QCM and ACIS measurements, we patterned $1 \times 1$ mm$^2$ to $3 \times 3$ mm$^2$ of Nafion film to 8 µC/cm$^2$ dose at 10 keV with 100 nm dwell-point spacing and a typical beam current of 240 pA via a 30 µm aperture using a Raith 150-Two EBL system. The typical write time was 6 minutes to 1 hour per sample (~0.1 cm$^2$/hour pattern rate). Figure 1d shows an optical microscopy image of a 230 nm Nafion film that was electron-beam processed with these settings. Our ellipsometry measurements required an area exceeding $5 \times 5$ mm$^2$ and for these we used a 60 µm aperture to give a beam current as high as 1 nA and a dwell point spacing of 300 nm, which gave a write speed of 0.4 cm$^2$/hour. The NR measurements require an even larger patterned area (~cm$^2$). We used a Raith Voyager EBL system to prepare the NR samples due to the superior patterning speed (0.86 cm$^2$/hour) compared to the Raith 150-Two system. The Raith Voyager has a fixed beam energy of 50 keV, which is much higher than the Raith 150-Two (10 keV). This reduces the electron-beam interaction volume within the Nafion,[32,33] and the dose was increased to 20 µC/cm$^2$ to compensate for this. A smaller secondary NR sample was prepared on our Raith 150-Two system (10 keV, 8 µC/cm$^2$) to confirm consistent behaviour between the 50 keV and 10 keV samples (see Supporting Information and Fig. S7).

Throughout the materials characterizations, electron-beam patterned films are compared with unpatterned films, and the latter are referred to as pristine films or control samples. For some measurements it was necessary to create a defined area of Nafion film on the control sample rather than an unconstrained film. In that case a spin-cleaning method was used, as follows. The desired area of the pristine Nafion film was covered with a thin piece of PDMS. The remaining Nafion was then washed away with a 10 s rinse with a 2-propanol jet while the sample was spinning at 3000 rpm using a spin-coater. The sample is then dried and the PDMS is removed. All control samples were placed under vacuum for 1-1.5 hours after spin coating, irrespective of whether they have defined area or are an unconstrained film. This was done to simulate the atmospheric conditions experienced by the electron-beam patterned samples during the patterning process.

We found some aging effects occur in Nafion films in the first day or two after spin-coating, affecting film conductivity. This occurs even when the samples are stored in an oxygen/water-free environment. This presents some stability issues for the measurements (see Supporting Information and Fig. S4) and is likely due to slow passive evaporation of solvent from the Nafion film. The aging behaviour was observed in both electron-beam patterned and pristine samples. To overcome this, all Nafion samples used for conductivity measurements were stored in a nitrogen glove box for at least three days prior to characterization, irrespective of whether they were patterned or pristine.

### 2.2.2. Chemical changes induced in Nafion by the electron-beam

We performed X-ray Photoelectron Spectroscopy (XPS) on four Nafion films on Si substrate with a native oxide surface to investigate chemical changes caused by electron-beam exposure. **Figure 2**(a-d) shows the carbon 1s and oxygen 1s spectra for two 230 nm films, one pristine (a,b) and one electron-beam patterned (c,d), with the raw data and peak fits presented.

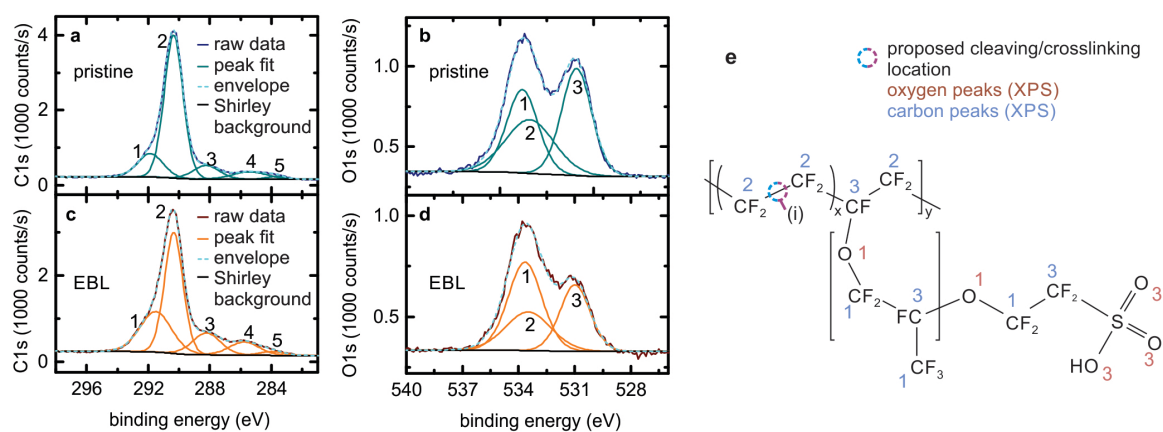

**Figure 2.** (a) C1s and (b) O1s spectra for a pristine Nafion film. (c) C1s and (d) O1s spectra for an electron-beam patterned Nafion film. (e) Chemical structure of Nafion-117 with numbering consistent with the XPS peak assignments.

The peak fits for the C1s and O1s spectra require five and three peaks, respectively. We start by considering the pristine films to connect each peak to the specific bonding chemistry. For the C1s spectra, Peak 1 includes difluorinated carbon bonded to ether oxygen and trifluorinated carbon. Peak 2 captures the difluorinated carbon on the polytetrafluoroethylene (PTFE) backbone. Peak 3 is assigned to monofluorinated carbon and the carbon attached to sulfonate groups. The bonds corresponding to these peaks are numbered accordingly in the chemical structure shown in Fig. 2e. The two remaining peaks are impurities with Peak 4 assigned to partially fluorinated carbon and Peak 5 to adventitious carbon. The relative areas for Peaks 1-3 are as expected for the Nafion-117 chemical structure (see Table S2). For the O1s spectra, Peak 1 corresponds to oxygen ether bonds within Nafion, Peak 2 to an impurity that may be $N_xO_y$ given the presence of nitrogen, and Peak 3 to sulfonated oxygen. The relative areas for Peaks 1 and 3 are also as expected (see Table S3). We have not shown the relevant sulfur 2p spectra in Figure 2 because these spectra do not show any significant change between pristine and electron-beam patterned samples (see Fig. S5).

We now move to the C1s and O1s spectra for the electron-beam patterned film (Fig. 2c/d), which show three key changes relative to the pristine film (full details in Tables S4 – S7). The first is a clear change in the relative area ratio of Peaks 1 and 3 in the O1s spectrum (Table S7). The increased area for Peak 1 suggests an increase in the relative amount of ether bonds in the film.[34] The C1s spectrum adds further insight to this. The character of Peak 1 in C1s changes significantly, with an increase in the binding energy, peak width, and peak area (Table S5). Additionally, the area of Peak 2 in C1s has decreased. This suggests that the electron-beam breaks the carbon-carbon bond in the $CF_2$–$CF_2$ chains on the PTFE backbone (see bond labelled (i) in Fig. 2e). Note that although there are $CF_2$–$CF_2$ bonds on the side-groups, the adjoining ether and sulfonate groups mean it does not contribute to Peak 2, i.e., changes in Peak 2 do not indicate cleavage at these side-group bonds. Considering the C1s and O1s spectra, a possible crosslinking mechanism is that free oxygen forms intermolecular ether bonds between pairs of the cleaved $CF_2$–$CF_2$ groups on nearby PTFE backbones. We believe this dominates over crosslinking between backbones and side-groups or between pairs of side-groups because the side-groups are rich in ether bonds, and cleavage on the side-groups would limit or even negate the increase in ether signature that we see in the O1s spectra. The increased chemical heterogeneity that the proposed backbone crosslinking mechanism provides would also explain the increased width for Peak 1 in the C1s spectrum.

The question then becomes where does the oxygen for these new ether bonds come from? Electron-beam patterning is done under high vacuum ($< 10^{-3}$ mbar), but given the water affinity and porosity of Nafion, residual water and/or trapped trace oxygen may contribute to this process, though detection of trace water specifically via XPS for confirmation appears infeasible considering the impurity (see SI for further discussion). Another possibility that we cannot rule out is that oxygen is scavenged from the sulfonate groups. However, since the sulfur spectra show no significant change in its chemistry overall, if scavenging of sulfonate groups is occurring then it is likely that the sulfonate is reoxidised by atmospheric oxygen once the sample is returned to ambient conditions after the electron-beam patterning process. However, given that the overall sulphur and oxygen content in the electron-beam processed film is lower than for the control sample, if this second mechanism is present, it is clearly leading to significant loss in sulfonate. It may be that upon cleaving the $SO_3$ group is reduced but a significant fraction is then reoxidised and evolves as a gas (e.g., $SO_2$) after exposure to atmosphere. On balance, the first mechanism is preferred, with the understanding that there is $SO_3$ loss.

### 2.2.3. Water absorption of electron-beam patterned films

We next investigate how the patterning process impacts the water absorption properties of Nafion, since these are crucial to its application as an ion-conducting material.[4] We used a quartz crystal microbalance[35] to measure the hydration number $\lambda$ of 230 nm and 30 nm Nafion films before and after electron-beam patterning. The hydration number is defined as the number of water molecules per sulfonic acid group $\lambda = n(H_2O)/n(SO_3)$. Note that $\lambda$ is calculated under the assumption that $\lambda = 0$ under vacuum conditions, i.e., there is no residual water.[36] Our methodology follows that developed by Shim et al.[36] Nafion films between $2 \times 2$ mm$^2$ and $3 \times 3$ mm$^2$ in size were prepared directly onto the Au electrode surface of a quartz crystal sensor, using the spin-cleaning technique mentioned earlier for the pristine films and EBL for the electron-beam processed films (see Fig. S6a). The Au electrode is convenient as it facilitates charge dissipation during electron-beam patterning. The QCM sensor was placed in a sealed chamber connected to the humidity control system[37] illustrated in Fig. S6b. Humidity control was executed by a process similar to that reported previously for melanin hydration studies.[38] In this method, we set the water activity $a_w = P/P_S$ via the pressure $P$ of water vapour in the chamber; $P_S$ is the saturation water vapour pressure at the given temperature. All measurements were carried out at room temperature, and we have corrected for measured fluctuations in room temperature to ensure best-possible correspondence between separate physical measurements.

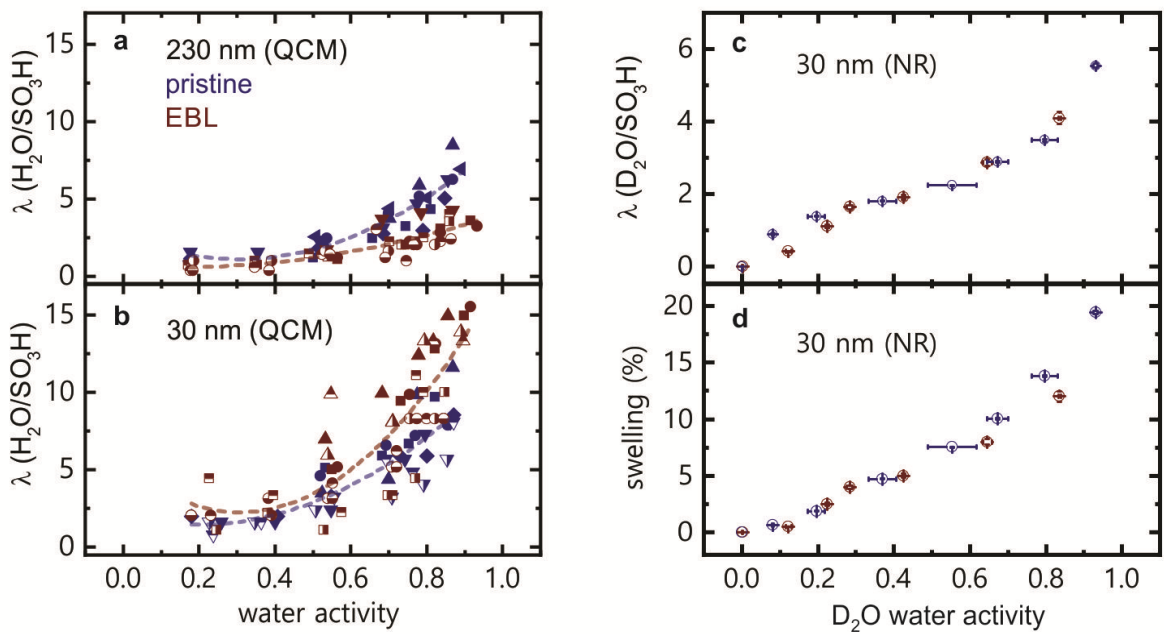

**Figure 3.** (a/b) Hydration number $\lambda$ vs water activity $a_w$ for (a) 230 nm and (b) 30 nm Nafion films. (c) $\lambda$ and (d) swelling vs $a_w$ obtained from neutron reflectometry (NR) measurements of 30 nm thick Nafion films with hydration by $D_2O$ rather than $H_2O$ for NR contrast. Data from pristine films is shown in blue (see also [48]) and electron-beam processed films in red. Different samples are represented by different shapes. Different half-shadings for a common shape are used to indicate separate measurements of the same sample. Water activity has been corrected for fluctuations in room temperature between measurements.

**Figures 3**a and b show $\lambda$ versus $a_w$ for pristine (blue) and electron-beam patterned (red) films. The dashed trend lines are second-order polynomial fits intended solely as a guide-to-the-eye. Both pristine and electron-beam patterned 230 nm films (Fig. 3a) show Type-II adsorption isotherm[39] behaviour, which is well known for Nafion membrane[23,40] and Nafion films with thickness exceeding 60 nm.[24,41–44] We observe lower water absorption for the electron-beam patterned film (red) than for the pristine film (blue). The hydration number $\lambda$ is approximately two times larger for the pristine film at $a_w = 0.9$. The reduced water uptake suggests a change in film structural properties that might be explained as follows. The film needs to swell to uptake water.[23,24,43] Electron-beam induced crosslinking may reduce the elasticity, i.e., increase Young's modulus, thereby limiting both the swelling and the ultimate water uptake capacity. This explanation is consistent with the AFM results in Section 2.1, the predicted behaviour of hydration number with Young's modulus,[45] and with studies of highly sulfonated aromatic polymers, developed as alternatives for Nafion, where chemical crosslinking has been shown to reduce water uptake at room temperature.[46]

The pristine 30 nm film behaves similarly to the 230 nm pristine film, albeit with a water uptake that is a factor of 1.2 higher at high $a_w$ (Fig. 3b). The most notable difference is that $\lambda$ for the electron-beam patterned samples is higher than for the corresponding pristine samples. There are several possible reasons for this. First, as seen in Fig. 1j, the patterning process leads to a reduction in film thickness by ~10 nm. The water absorption in thin films <60 nm is very sensitive to changes in thickness with λ having been shown to increase rapidly as the thickness decreases.[47] It is therefore possible that any structural changes that may cause stiffening of the film are counteracted or overcome by an increase in water uptake due to a reduction in film thickness. Note that we were unable to obtain precise thickness measurements of the individual Nafion films on the QCM crystals due to the high surface roughness of the QCM sensors. Second, the Nafion film near the substrate interface may interact differently with the electron-beam compared to the bulk layer that likely dominates the adsorption behaviour of the 230 nm films. Overall, the influence of film thickness on water uptake is a complicated issue, with both decreasing[36,47,48] and increasing[22,24,42,43] dependencies observed in previous experiments. The nature of the surface of the underlying substrate also plays a particularly vital role in the measured water uptake for films with thickness well below 100 nm.[36,43] The novelty of our electron-beam patterned material[16] enabled us to obtain neutron reflectometry time to further explore the water uptake behaviour for the 30 nm films reported in Fig. 3b. An advantage of using NR here is that we can measure on the same device-quality $SiO_2$-on-Si substrates that we use for our Nafion inverter devices for bioelectronics applications.[16]

A pristine 30 nm Nafion film control sample for NR was spin-coated on a 2-inch silicon wafer and measured 'as is' on the INTER instrument at the Isis Neutron and Muon Source at the Rutherford Appleton Laboratory. These results were published elsewhere[49] but the adsorption result is used here (Fig. 3c (blue)) to enable comparison. The measurement at INTER used a custom-built water vapour delivery system with manually operated valves; the uncertainty in $a_w$ in Figs. 3c/d is thus correspondingly slightly larger for this measurement. The experiments were performed using $D_2O$ for hydration over $H_2O$ in the NR water uptake measurement. The reasons for this are described in Ref. 48 but can be summarised by understanding that the higher scattering length density (SLD) provided by $D_2O$ versus $H_2O$ allows for greater accuracy in bulk water uptake, at the cost of more easily identifying water distribution within the film.[49,50] The NR measurements of a 30 nm electron-beam patterned

Nafion film (Fig. S11 and S12) were performed on the Platypus instrument[51] at the OPAL reactor at the Australian Nuclear Science and Technology Organisation (ANSTO). At ANSTO, water activity control was achieved using a modified Hiden Isochema XCS system. This system follows the same operating principle as the apparatus in Fig. S6b except that the valves are computer controlled, allowing more continuous adjustment, and thereby reduced drift in $a_w$ for the electron-beam patterned data in Figs. 3c/d. The samples used for the NR measurements discussed below were obtained from samples electron-beam patterned at 50 keV rather than 10 keV to enable sufficient patterned area for the study. A study to demonstrate that the measured scattering length density is not substantially changed by the beam energy used for patterning is presented in the Supporting Information (see Fig. S7).

Figure 3c shows the measured λ versus $a_w$ extracted from an analysis of our NR data over the full range $0 < a_w < 1$ (see Methods for details). The adsorption isotherm for both films is Type-II,[39] following the form expected for Nafion membrane[23,40] and thin-film.[24,41–44] It is also qualitatively consistent with our own QCM data in Figure 3b, albeit with a quantitatively lower λ. We completed the NR study using films with identical target thickness and patterned one but not the other to obtain a meaningful control for the effect of electron-beam patterning. However, the NR study indicated a thickness for the electron-beam patterned film of 20.1 nm at vacuum whilst the unpatterned film had a thickness of 31.9 nm at vacuum. We obtained 22.4 nm and 36.4 nm respectively at high $a_w$ (see swelling data in Figure 3d). This is consistent with the ellipsometry data shown in Fig. 1j. The thickness difference motivated the AFM studies in Section 2.1 and is also consistent with the findings there. For a better quantitative understanding of how λ varies with electron-beam patterning, the electron-beam patterned sample should be compared to a pristine film with the same thickness as the patterned sample after the development process. This ideally would require a much more detailed NR study of patterned and unpatterned films over a wide film thickness range. However, there are some key counter points. The first is that the swelling behaviour of the NR tested films are very similar (Fig. 3d), which suggest that a good comparison can be drawn despite the difficulties mentioned earlier. Furthermore, some of the above differences in sorption behaviour can be due to methodology, i.e., QCM versus NR. NR is fundamentally based upon atomic composition and density whereas QCM is a frequency-based technique. As such, NR may be preferable, but further investigation would be needed to decide on the optimal technique for investigating Nafion thin films.

Ultimately, the qualitative aspects of our study shed sufficient light to provide a starting model for the effect of the electron-beam on water uptake in the 30 nm films to motivate future studies. Figure 3d shows the proportional swelling of the film measured by NR versus $a_w$, which suggests the near-substrate Nafion structure, and its mechanical properties, are not massively changed by the electron-beam. Interactions with the substrate exert significant influence on the near-substrate structure of a Nafion film even for thicker bulk-like films.[23] These interactions cause ordering of the water transport domains parallel to the interface[52,53] near the substrate, producing a water-rich inter-lamellar layer in the ~50 nm closest to the substrate.[21] The bulk-like layer that sits above this in thicker films shows a markedly different structure with crystallite formation and a higher degree of phase separation than the interfacial region.[23,47] The morphological differences between these layers may lead to very different outcomes regarding electron-beam-induced crosslinking, which may in turn drive the differences in water uptake behaviour between the 30 nm and 230 nm electron-beam patterned films in Figs. 3a and b. We also cannot rule out that the electron-beam may enhance the hydrophobic skin layer proposed by Kongkanand,[42] further limiting water uptake in the electron-beam patterned 230 nm film. The absence of these effects in the 30 nm film is consistent with the findings of Kusoglu et al.[54] where the lack of crystallite formation and lower phase separation mean that the mechanical properties became insufficient to limit hydration.

## 2.2.4. Ionic Conductivity of electron-beam patterned films

We now turn to the characterization of the ionic conductivity of electron-beam patterned and pristine Nafion films, which were performed on interdigitated electrode (IDE) chips with one of two designs: 1) twenty-six 20 μm wide electrodes with 30 μm spacing and 1 × 1 mm² active area, or 2) fifty 16 μm wide electrodes with 30 μm spacing and 2 × 2 mm² active area. A typical electron-beam patterned Nafion film on an IDE chip with Design 1 is shown in Fig. S6c; the two white-dashed lines indicate the film corners. The two designs were selected to best match each film's electrical properties to the ACIS instrumentation to optimise measurement accuracy. We measured the AC film ionic conductivity σ versus $a_w$ using the apparatus in Fig. S6d (see Methods and Supporting Information for details). Impedance spectroscopy was performed using a 10 mV excitation over the frequency range 1 Hz $< f <$ 100 kHz and the resulting spectra were fitted using models developed by Paul et al.[55] for Nafion thin films (see Fig. S8). The model consists of a series resistance, two constant-phase elements, and the film resistance $R_f$, which we extract and then convert to conductivity σ using[55]:

$$\sigma = \frac{d}{R_f\, l\, (N-1)\, t} \qquad (1)$$

where $d$ is the IDE electrode spacing, $l$ is the electrode length, $N$ is the number of electrodes in the array, and $t$ is the film thickness. One challenge here is that the electrode topography and roughness make it difficult to accurately measure the actual film thickness on the IDE chips. Instead, we based our thickness estimates on ellipsometry measurements of similar films made on $SiO_2$-on-Si substrate. The resulting film thicknesses used for obtaining the conductivity σ are 230 ± 23 nm (pristine) and 220 ± 40 nm (electron-beam patterned) for the thick films and 33 ± 3 nm (pristine) and 20 ± 4 nm (electron-beam patterned) for the thin films. Another challenge is the relatively slow equilibration of $a_w$, which we account for with the hydration and stabilization protocol discussed in the Supporting Information (see Fig. S9).

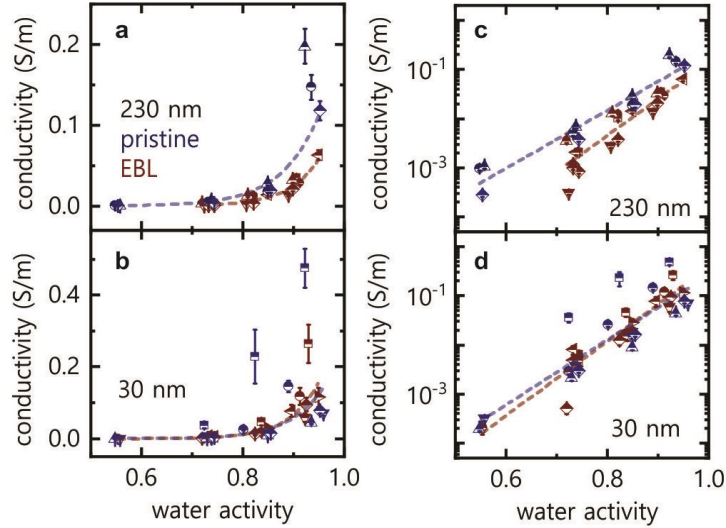

**Figure 4.** (a-d) Conductivity σ versus water activity $a_w$ for (a,c) ~230 nm and (b,d) ~30 nm Nafion films with electron-beam patterned film data in red and pristine film data in blue. Conductivity data is presented on a linear scale in a/b and a log scale in c/d. The dashed lines in (a-d) are guides to the eye.

Figures 4a-d show σ versus $a_w$ for electron-beam patterned (red) and pristine (blue) films with target thickness 230 nm (a/c) and 30 nm (b/d), with σ data presented on a linear scale (a/b) and on a log scale (c/d) to provide better clarity of the trends. The trend lines in (a-d) are purely intended as guides to the eye. The data in (a-d) plotted in a single panel, along with the underlying sheet resistance data, are given in Fig. S10. The scatter in conductivity primarily reflects sample-to-sample variations. Despite the scatter, we see robust trends across the various films measured. Starting with the 230 nm films, the conductivity of both pristine and electron-beam processed films increases approximately exponentially for $a_w$ > 0.5. This is a well-known behaviour for Nafion films.[55,56] Our conductivity values (0.001 – 0.1 S/m) are at least an order of magnitude lower than those obtained by Siroma *et al*.[56] (0.1 – 3 S/m) or Paul *et al*.[55] (0.5 – 5 S/m) for Nafion films. We are unclear on the exact cause, but note that variations in thickness, aging, deposition and processing can produce differences in conductivity at order of magnitude level in Nafion films.[23] Even manufactured Nafion membrane shows conductance variability at order-of-magnitude level between studies.[54] The key finding in this work is that the conductivity of the electron-beam patterned films is lower than the pristine films, but only by an order-of-magnitude at most. The conductivity σ is approximately 2-3 times lower at high

$a_w$ for the electron-beam patterned films than for the pristine films (see Fig. 4a). This is consistent with the reduction in water uptake in the 230 nm electron-beam patterned samples observed in the QCM measurements (Fig. 3a). This indicates that the reduction in σ is likely to be largely driven by reduced water uptake resulting from the increased material stiffness due to crosslinking as discussed in Section 2.2.3. We note that studies by Thankamony *et al.*[46] also showed that chemical crosslinking could reduce the room temperature conductivity by about an order of magnitude in highly sulfonated aromatic polymer films.

The 30 nm films in Fig. 4b/d show similar behaviour of exponentially increasing σ with increasing $a_w$, as expected.[55] The conductivity for the 30 nm films (0.001 – 0.1 S/m) is generally slightly lower than for the 230 nm films (0.0001 – 0.1 S/m), consistent with previous findings that the ionic conductivity decreases with thickness.[47,55,57,58] The reduction in conductivity with thickness is smallest near full hydration and increases with reduced $a_w$, consistent with studies of σ versus $a_w$ with film thickness by Paul *et al.*[55] We do not observe a significant difference in σ between electron-beam patterned and pristine 30 nm films. This is consistent with thickness dependent measurements on pristine Nafion films where a significant increase in water absorption with decreasing film thickness is observed in QCM measurements.[47] This, however, is not associated with an increase in conductivity.[47,55,57,58]

In summary, a key observation of our work is that the electron-beam patterning affects the ionic conductivity of Nafion at approximately order of magnitude level but the mechanism for this likely differs between thin films and thick films. The changes we see in the thick films appear to be driven by changes in material properties with electron-beam exposure, specifically, that crosslinking limits swelling, which in turn limits water uptake and thereby ionic conductivity. In contrast, the conductivity changes in the thin films are most likely driven instead by the reduction in film thickness resulting from the electron-beam process. This reduction also occurs for the thick films, but it is proportionally far more significant for the thin films, and in this thickness range, the water uptake measured by QCM is also extremely sensitive to film thickness.[47] Although this might immediately suggest a strong change in ionic conductivity for the thin films, previous studies have also found little change in conductivity despite the strong sensitivity of water uptake to thickness for Nafion thin films.[47,55,57,58] As such, the electron-beam patterned material does not appear to give a radical departure in behaviour from pristine films. Some substantial structural studies as a function of thickness would be needed to definitively demonstrate these mechanisms more conclusively, and we encourage this as a focus of future work on electron-beam patterned Nafion.

## 3. Conclusions

We have studied the effect of electron-beam patterning of a Nafion thin film on water uptake and ionic conductivity using a combination of XPS, QCM, NR and ACIS studies for two different target Nafion film thicknesses: 230 nm and 30 nm. The thicker films have a thickness equivalent to that used in our earlier work demonstrating the patterning of nanoscale structures in Nafion by electron-beam lithography and provide insight into how electron-beam processing influences Nafion in the bulk-like limit. The thinner films provide improved resolution under patterning but are also of interest because the behaviour of Nafion films changes markedly for thicknesses well below 100 nm.

The materials characterizations for this study require total patterned areas (mm$^2$ – cm$^2$) that are far larger than those required for nanoscale device applications (μm$^2$). To facilitate this, we optimized electron-beam patterning of Nafion for large-scale exposures enabling patterning speed as high as ~0.86 cm$^2$/hr using standard commercial electron beam lithography systems. We patterned areas up to 7 cm$^2$. Combining our optimisation for large areas with the parameters for nanoscale resolution offers a path to scalable electron-beam patterning of Nafion films for electronic applications. Additionally, we found that certain exposure parameter configurations enabled non-flat, textured Nafion films to be produced that might also have applications potential.

XPS studies of pristine and electron-beam patterned Nafion films show changes in the O1s and C1s spectra that point strongly to an increase in ether bond percentage after electron-beam patterning, whilst the S2p spectra is largely unaffected. The data suggests a mechanism where the C–C bonds in the $CF_2$–$CF_2$ moiety of the PTFE back-bone are cleaved by the electron-beam, and then react with oxygen to form intermolecular ether bonds that crosslink the film and provide contrast in the developer solution. The necessary oxygen likely comes from residual water or atmospheric oxygen trapped in the nanoporous film structure. We cannot rule out some contribution from cleavage of oxygen at side-group sulfonate or ether groups, but if this is occurring, these likely re-oxygenate once the sample is returned to ambient conditions, otherwise the unaffected S2p spectra and higher ether bond percentages cannot be explained.

QCM studies show that electron-beam patterning produces different outcomes depending on the thickness of the film. All films show Type-II adsorption isotherm behaviour, with the hydration increasing at an ever-faster rate with increasing $a_w$. This is expected based on past studies of Nafion membrane[23,40] and films.[24,41–44] A hydration of $\lambda \sim 8$ is obtained as $a_w$ approaches 1 for both thicknesses of pristine film. The electron-beam patterned 230 nm films show lower adsorption giving $\lambda \sim 4$ as $a_w$ approaches 1. This is a reduction in hydration relative to the pristine equivalent by approximately a factor of 2. We studied the 30 nm electron-beam patterned films by two methods: QCM and NR. The QCM data suggests a higher $\lambda$ relative to the pristine films with $\lambda$ as high as 15 as $a_w$ approaches 1. However, the NR data instead showed no significant difference in $\lambda$ between electron-beam patterned and pristine films. We suspect this discrepancy arises from fundamental differences between the two techniques and possibly also the different substrate materials involved, i.e., Au for QCM and $SiO_2$ for NR, which have been shown to produce differences in water uptake in sub-60nm Nafion films in previous work.[22,42,47] We suggest the reduced water absorption in the 230 nm film is driven by electron-beam induced crosslinking, which stiffens the material and inhibits the swelling that is essential to facilitating water uptake. We found that film thickness is reduced by the electron-beam patterning process. Ellipsometry, NR and AFM measurements indicate a typical thickness reduction of approximately 10 nm. The water absorption of Nafion films with thickness less than ~60 nm has been shown to increase drastically with decreasing thickness.[47] Thus, we suggest that increased water absorption caused by film thickness reduction compensates any absorption inhibiting effects of the crosslinking in the 30 nm films.

Finally, we used ACIS studies to investigate the effect of electron-beam patterning on the ionic conductivity $\sigma$. Electron-beam patterning leads to a reduction in conductivity by a factor of 2-3 at high $a_w$ for the 230 nm films, which is consistent with the reduction of water uptake observed in the QCM measurements. For the 30 nm films, we see no significant change in $\sigma$, which is consistent with the NR measurements where there is correspondingly no significant change in water uptake. In the context of the QCM data, our findings are consistent with thickness dependent measurements on pristine Nafion films where there is a significant increase in water absorption with decreasing film thickness, but no associated increase in conductivity is observed.[47,55,57,58] The changes we see in the thick films appear to be driven by changes in material properties with electron-beam exposure, specifically, that crosslinking limits swelling, which in turn limits water uptake and thereby the ionic conductivity. In contrast, the changes in the thin films are most likely driven instead by the reduction in film thickness resulting from the electron-beam process.

Ultimately more substantial structural studies as a function of thickness over a range of thicknesses and electron-beam patterning parameters would be required to form a more definitive picture of the full effect of electron-beam interactions on Nafion. However, our work points to electron-beam patterning having a significant thickness-dependent effect on water uptake and a slight detrimental effect on ionic conductivity for Nafion films. These results provide a useful basis to inform future use of electron-beam patterned Nafion in future device applications and offer an interesting entry point for further studies of the materials physics of Nafion films more broadly.

**4. Methods Section**

*Nafion film preparation*: All substrate surfaces were cleaned with acetone and 2-propanol, with sonication where possible, and then dried with nitrogen gas prior to Nafion deposition. The films with 230 nm target thickness were spin-coated at 3000 rpm for 30 s from the undiluted stock solution. The films with 30 nm target thickness were spin-coated from a 1:4 mixture of stock solution in ethanol[22] at 3500 rpm for 60 s. The stock solution was a 5% suspension of Nafion-117 in lower aliphatic alcohols and water (Aldrich 70160). Films used as control samples, i.e., without electron-beam patterning, were placed under vacuum for 60-90 minutes to emulate the conditions experienced by the electron-beam patterned samples during the electron-beam patterning process.

*Electron-beam patterning*: Samples for XPS, QCM, and ACIS were patterned with a 10 keV electron beam with a current of 240 pA and dwell point spacing of 100 nm to an area dose of 8 μC/cm$^2$ using the Raith 150-Two system at UNSW. Samples for ellipsometry were exposed at 10 keV beam energy, 1 nA beam current, 500 nm dwell point spacing and 8 μC/cm$^2$ dose using a Raith 150-Two system. The primary 23 × 30 mm$^2$ NR sample was patterned at 50 keV beam energy, 1 nA beam current, 500 nm dwell point spacing and 20 μC/cm$^2$ dose using the Raith Voyager system at Lund University. A smaller back-up NR sample (see Supporting Information) was exposed at 10 keV beam energy, 1 nA beam current, 500 nm dwell point spacings and 8 μC/cm$^2$ dose using a Raith 150-Two system. All samples were developed in a 1:1 mixture of 2-propanol and acetone for 60 s.

*X-ray photoelectron spectroscopy*: XPS was performed on a Kratos Axis Supra using a 225 W Al Kα x-ray source with 15 mA emission current and a quartz crystal monochromator with 500 mm Rowland circle. The x-rays typically illuminate a 300 × 700 μm$^2$ area with sampling depth limited by the electron mean free path in this energy range (< 10 nm). High-resolution spectra were collected with a pass energy of 40 eV, 0.1 eV step size, 1 s counting dwell time at each step, and up to four sweeps depending on the signal-to-noise ratio. We used a hybrid lens setting where both electrostatic and magnetic immersion lenses are deployed to collect the photoelectrons. An integral Kratos charge neutralizer was used as an electron source to eliminate differential charging. Peak analysis of XPS spectra was performed in the CasaXPS software (2.3.17dev6.4k) using the Kratos sensitivity factor library. A Shirley background was used and mixed Gaussian-Lorentzian peaks with 30% Lorentzian character (GL(30)) were fitted to the raw data. After data fitting the energy axis of all spectra was charge calibrated to the C–C component in the C1s fit at 284.8 eV. Further details are given in the supporting information.

*Atomic force microscopy*: AFM was performed using a JPK NanoWizard II (Bruker) scanning probe system in AC mode. Bruker NCHV tips were used with a drive frequency of 300 kHz. Image analysis was performed using Gwyddion software.[59]

*Humidity control*: Humidity control followed a process used previously for melanin hydration studies.[38] In this method we set the water activity $a_w = P/P_S$ via the pressure $P$ of water vapour in the chamber where $P_S$ is the saturation water vapour pressure at the given temperature. All measurements were carried out at room temperature, and we have corrected for measured fluctuations in room temperature to ensure best-possible correspondence in $a_w$ for separate physical measurements across different apparatus/characterization methods. The following process was used to set a humidity for measurement (see Fig. S6b,d). First, the sample chamber was evacuated via Valve 1 (V1) with Valve 2 (V2) closed. The chamber was isolated from the pump by closing V1 after 60 min of pumping and V2 was then carefully opened to bleed water vapour from a vial containing degassed deionised water into the chamber. V2 was closed once the desired $P$ was reached. Thereafter $P$ slowly decreases because water gradually condenses onto the chamber walls and is absorbed by the sample. As a result, we spent the next 30 minutes repeatedly opening V2 briefly (~1 min) to replenish the water vapour in the chamber until $P$ remained stable at the desired value. The sample was then left to equilibrate for 15-30 min prior to measurement. We found that 15-30 min of wait time after manually stabilising $P$ was sufficient for stable and repeatable measurement conditions to be achieved (see Supporting Information and Fig. S9). The process above was repeated for each pressure point, i.e., the chamber was evacuated between measurements at different $a_w$ values, to ensure consistent equilibration between data points and avoid hysteresis due to over-condensation of water inside the chamber.

*Quartz crystal microbalance:* We used a Inficon Q-Pod quartz crystal microbalance with 14 mm diameter gold-coated 6 MHz quartz crystal sensors (Inficon 750-1000-G10). Nafion was spin-coated directly on the quartz crystal sensor and then patterned in one of two ways. For electron-beam patterned samples, an area between $2 \times 2$ mm$^2$ and $3 \times 3$ mm$^2$ were exposed and developed using the electron-beam lithography parameters specified earlier. For the control samples, an area between $2 \times 2$ mm$^2$ and $4 \times 4$ mm$^2$ was prepared using a spin-cleaning technique as follows. The desired area of pristine Nafion film was covered with an appropriately sized piece of PDMS film. The remaining Nafion was then washed away with a 10 s rinse with a 2-propanol jet while the sample was spinning at 3000 rpm using a spin-coater. For each sensor, the crystal resonance frequency was measured prior to Nafion deposition, and then after patterning, with the Nafion film under vacuum and at various set $a_w$ values. The hydration number $\lambda$ was calculated using $\lambda = (m_{H2O} \times 1000) / (M_{H2O} \times m_d \times IEC)$ where $m_{H2O}$ is the mass of absorbed water, $m_d$ is the mass of dry Nafion measured under vacuum, $M_{H2O}$ is the molar mass of water and $IEC = 0.909$ mmol/g is the ion exchange capacity for Nafion.[36] The dry Nafion mass $m_d$ was obtained from the Sauerbrey equation:[35] $f_{bare} - f_{dn} = (-2 \times m_d \times f_{bare}^2) / [A \times (\mu \times \rho_q)^{0.5}]$, where $f_{bare}$ is the bare crystal frequency, $f_{dn}$ is the crystal frequency with dry Nafion under vacuum, $A = 3.22 \times 10^{-5}$ m$^2$ is the crystal area, $\mu = 2.95 \times 10^{11}$ dyn/cm$^2$ is the crystal shear modulus, and $\rho_q = 2.65$ g/cm$^3$ is the density of quartz. Under hydrated conditions, the crystal frequency was monitored for 10-15 min and averaged to obtain the mass of water absorbed by the Nafion film.

*Neutron reflectometry*: Specular NR data were acquired on the *Platypus* time-of-flight neutron reflectometer at ANSTO.[51] A wavelength spectrum of 0.28 to 1.8 nm was used, at angles of incidence of 0.65, 2.0 and 4.5 degrees, corresponding to a *Q* range of 0.08 to 3.5 nm$^{-1}$. The overall instrumental resolution is independent of *Q*, *dQ*/*Q* ~ 8.4%. In the NR experiment the partial pressure of water was computer controlled by a Hiden Isochema XCS system, with the sample mounted in a stainless-steel vacuum vessel.

NR data were fitted and analysed using RasCAL2019 in MATLAB R2020a. The sample was modelled as a single layer of Nafion with uniform thickness, roughness, and scattering length density (SLD). More complicated models were examined (multiple slabs) but did not result in a significant improvement in the fit quality. A native oxide layer was included on the silicon substrate. Fitting was performed on each dataset with all $a_w$ profiles modelled simultaneously to improve fit confidence in the parameters for the substrate (SiO$_2$ thickness, etc). Confidence intervals for the fit parameters were generated via a bootstrapping method, using 250 iterations and random parameter initialization. The resulting empirical cumulative distribution functions were interrogated to obtain a 99% confidence interval for the resulting film thickness and the integrated area under the SLD profile. Integrations were performed using limits calculated to avoid the inclusion of the substrate contributions to SLD. The specific data that NR provided are the water uptake and changes in Nafion thickness. The thickness change/swelling is calculated from the proportional change in Nafion film thickness between the thickness of the Nafion at the given $a_w$ and the thickness at vacuum. The water uptake is calculated from the following equation:

$$\lambda = \frac{b_i}{b_w} \frac{\int SLD_{Tot}^{highRH} - \int SLD^{lowRH}}{\int SLD_{Tot}^{lowRH}} = \frac{b_i}{b_w} \left( \frac{\int SLD_{Tot}^{highRH}}{\int SLD_{Tot}^{lowRH}} - 1 \right) \quad (2)$$

where $b_i$ and $b_w$ are literature values[60] for the sum of the bound coherent neutron scattering lengths of each of the nuclei in the formula units for the Nafion and water, respectively (see Table S8). A more complete derivation of Eqn. 2 is given in the Supporting Information. The integrals are the total integrated areas under the SLD profile for the high $a_w$ and low $a_w$ (i.e., vacuum) measurements. We calculate $\lambda$ under the assumption that $\lambda = 0$ for Nafion films at vacuum. Further details about the water uptake calculations are provided in the Supporting Information.

*Ionic conductivity measurements*: Custom IDE chips with $1 \times 1$ mm$^2$ and $2 \times 2$ mm$^2$ IDE areas were prepared by photolithography on glass. The 1 mm$^2$ chips feature 26 finger electrodes, each 1000 μm in length and 20 μm in width with a 30 μm gap between adjacent finger electrodes. The 4 mm$^2$ chips have 50 finger electrodes, each 2000 μm in length, 16 μm in width with a 30 μm gap between adjacent finger electrodes. Use of the larger 4 mm$^2$ IDE enabled a more accurate measurement of high resistivity samples, e.g., the 30 nm films at low to moderate $a_w$ because this geometry gives an overall smaller resistance relative to the 1 mm$^2$ IDE geometry for a film with an equivalent sheet resistance. For pristine samples, we spin-coated Nafion and patterned it by spin-cleaning, as described earlier, to produce films slightly larger than the electrode area ($1.1 \times 1.1$ mm$^2$ or $2.1 \times 2.1$ mm$^2$) to ensure complete electrode coverage irrespective of any small alignment error of the PDMS mask. The electron-beam patterned samples were patterned instead by EBL as described earlier. For the ACIS measurements, a Stanford SR830 lock-in amplifier was used to apply a sinusoidal voltage with 10 mV amplitude and frequency 1 Hz $< f <$ 100 kHz to one electrode of the IDE chip. The other electrode was connected via a Femto DLPCA-200 current-to-voltage amplifier to the lock-in amplifier voltage input to measure the response. The humidity control system is the same used for the QCM measurements (see Fig. S6d). The Nafion film resistance $R_f$ was obtained by fitting the data in the Nova 2.1 software (Metrohm Autolab B.V.) using an equivalent circuit model reported by

Paul *et al.*[55] (see Fig. S8) for details. The ionic conductivity σ of Nafion was calculated using

$\sigma = d/(R_f l(N-1)t)$ where $d$ is the distance between two adjacent IDE fingers, $l$ is the length of

an individual finger electrode, $N$ is the number of fingers, and $t$ is the Nafion film thickness.

**Supporting Information**

Fig. S1: Atomic force microscopy scans of Nafion patterned with focussed & defocussed beams with differing dwell point spacings.

Fig. S2: Illustration of the Nafion patterning process with high area dose.

Fig. S3: AFM image of four adjacent squares patterned with differing electron-beam dose.

Fig. S4: Conductivity vs days since fabrication demonstrating the aging of our films.

Fig. S5: Additional XPS data including F1s, S2p and N1s spectra.

Table S1: Atomic percentages for the atoms from the Nafion samples.

Table S2: Peak parameters for the C1s data from the control sample.

Table S3: Peak parameters for the O1s data from the control sample.

Table S4: Peak parameters for the C1s data from the electron-beam patterned sample.

Table S5: Change in peak parameters for the C1s atoms between pristine and electron-beam patterned Nafion.

Table S6: Peak parameters for the O1s atoms in the electron-beam patterned sample.

Table S7: Change in peak parameters for the O1s atoms between pristine and electron-beam patterned Nafion.

Fig. S6: Apparatus used for the QCM and ACIS measurements.

Table S8: Values for the neutron scattering lengths in each material.

Fig. S7: NR measurements of hydration number and relative swelling for 30nm Nafion films patterned at beam energies of 10 kV and 50 kV.

Fig. S8: Equivalent circuit model and representative Nyquist plot for the ACIS measurements.

Table S9: Fitted values and estimated error for the various circuit elements in Fig. S8.

Fig. S9: Data illustrating the hydration procedure needed to obtain stable measurement conditions.

Fig. S10: Additional sheet conductance data for the Nafion films.

Fig. S11: Neutron reflectivity profiles for the Nafion films.

Fig. S12: Scattering length density profiles from fits to the NR data.

**Acknowledgements**

This work was funded by the Australian Research Council (ARC) under DP170104024 and DP170102552, the Welsh European Funding Office (European Regional Development Fund) through the Sêr Cymru II Program. P.M. is a Sêr Cymru Research Chair and an Honorary Professor at the University of Queensland. A.B.M. contribution was under the Sêr Cymru II fellowship and the results incorporated in this work have received funding from the European Union's Horizon 2020 research and innovation program under the Marie Skłodowska-Curie grant agreement No 663830. The work was performed in part using the NSW and Queensland nodes of the Australian National Fabrication Facility (ANFF) and the Electron Microscope Unit (EMU) within the Mark Wainwright Analytical Centre (MWAC) at UNSW. The electron-beam patterning was performed in part at Lund NanoLab at Lund University. We acknowledge the provision of beamtime by ANSTO under proposal number P8773.

**Table of Contents Figure**

We explore how electron-beam patterning affects the water uptake and ionic conductivity of Nafion films on silicon substrates motivated by possible applications of this material in bioelectronics and neuromorphic computing. Our key result is that the reduction in ionic conductivity arising from electron-beam-generated cross-linking is less than an order of magnitude.

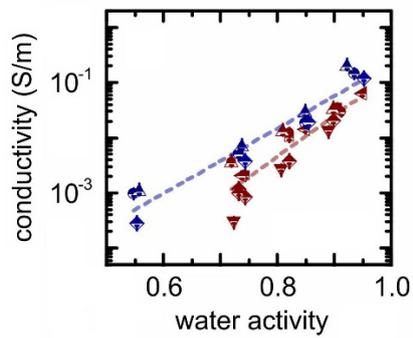